\documentclass[twocolumn,prl,superscriptaddress,showpacs,letter,aps,10pt,floatfix]{revtex4-2}

\usepackage{times}
\usepackage{graphicx}
\usepackage{dcolumn}
\usepackage{bm}
\usepackage{mathtools}
\usepackage[english]{babel}
\usepackage{physics}
\usepackage{enumitem}
\usepackage{mathrsfs}
\usepackage{xfrac}
\usepackage{xr-hyper}
\usepackage{footmisc}
\usepackage{makecell}
\usepackage{url}
\usepackage{ifsym}
\usepackage{footmisc}
\usepackage{makecell}
\usepackage{url}
\usepackage{ifsym}
\usepackage{bbold}
\usepackage{gensymb}
\usepackage{multirow}
\usepackage{array}
\usepackage{longtable}
\usepackage{siunitx}
\usepackage{nicefrac}
\usepackage[dvipsnames]{xcolor}
\usepackage{ulem}
\usepackage{comment}
\usepackage{amsmath}
\usepackage{hyperref}
\usepackage{hypcap}

\newcommand{\ci}{\mathrm{i}}
\newcommand{\ee}{\mathrm{e}}

\newcommand{\tp}{\mathrm{p}}
\newcommand{\tB}{\mathrm{B}}
\newcommand{\tpr}{\mathrm{pr}}

\newcommand{\trr}{\mathrm{res}}

\newcommand{\bs}[1]{\boldsymbol{#1}}

\DeclareMathOperator{\di}{d\!}
\AtBeginDocument{\RenewCommandCopy\qty\SI}

\newcommand{\beq}{\begin{equation}}
\newcommand{\eeq}{\end{equation}}

\newcommand{\bea}{\begin{eqnarray}}
\newcommand{\eea}{\end{eqnarray}}

\newcommand{\mathcheck}[1]{\color{dogwoodrose}{#1}}
\newcommand{\tcp}[1]{\mathcheck{\tilde{c}_\tp}}

\begin{document}

\title{Multiply quantized vortex spectroscopy in a quantum fluid of light}

\author{Killian Guerrero}\affiliation{Laboratoire Kastler Brossel, Sorbonne Universit\'{e}, CNRS, ENS-Universit\'{e} PSL, Coll\`{e}ge de France, Paris 75005, France}
\author{K\'evin Falque}\affiliation{Laboratoire Kastler Brossel, Sorbonne Universit\'{e}, CNRS, ENS-Universit\'{e} PSL, Coll\`{e}ge de France, Paris 75005, France}
\author{Elisabeth Giacobino}\affiliation{Laboratoire Kastler Brossel, Sorbonne Universit\'{e}, CNRS, ENS-Universit\'{e} PSL, Coll\`{e}ge de France, Paris 75005, France}
\author{Alberto Bramati}\email{alberto.bramati@lkb.upmc.fr}\affiliation{Laboratoire Kastler Brossel, Sorbonne Universit\'{e}, CNRS, ENS-Universit\'{e} PSL, Coll\`{e}ge de France, Paris 75005, France}
\author{Maxime J. Jacquet}\email{maxime.jacquet@lkb.upmc.fr}\affiliation{Laboratoire Kastler Brossel, Sorbonne Universit\'{e}, CNRS, ENS-Universit\'{e} PSL, Coll\`{e}ge de France, Paris 75005, France}

\begin{abstract}
The formation of quantized vortices is a unifying feature of quantum mechanical systems, making it a premier means for fundamental and comparative studies of different quantum fluids. Being excited states of motion, vortices are normally unstable towards relaxation into lower energy states. However, here we exploit the driven-dissipative nature of polaritonic fluids of light to create stationary, multiply charged vortices. We measure the spectrum of collective excitations and observe negative energy modes at the core and positive energy modes at large radii. Their coexistence at the same frequency normally causes the dynamical instability, but here intrinsic losses stabilize the system, allowing for phase pinning by the pump on macroscopic scales. We observe generic features of quantized vortices in quantum fluids and other rotating geometries like astrophysical compact objects, opening the way to the study of generic amplification phenomena.
\end{abstract}

\maketitle
Quantum fluids exhibit quantized vortices, topological defects around which the circulation of velocity is quantized in units of $h/m^\star$ with $h$ the Planck constant and $m^\star$ the mass of a fluid particle~\cite{feynman_chapter_1955}.
Experiments~\cite{engels_giant_2003,shin_mqvinsta_2004,Isoshima_mqvinsta_2007,okano_splitting_2007,dominici_interactions_2018} show that a multiply quantized vortex (MQV), a vortex with a winding number $C>1$, will spontaneously decay into a cluster of $C=1$ singly quantized vortices (SQV) to reduce the total energy of the system~\cite{barenghi_primer_2016}.

The dynamical instability of MQVs is now understood in terms of the build-up of excitations within an ``ergoregion''~\cite{giacomelli_ergoregion_2020}---a finite-radius region localized at the MQV core where collective excitations can acquire negative energy (work done on these excitations reduces locally the energy of the system~\cite{castin_bose-einstein_2001}).
If this ergoregion is surrounded by a region that supports positive energy modes, spontaneous and simultaneous amplification of negative and positive energy excitations can occur, conserving the total energy of the system. This is a field effect generic to rotating geometries called rotational superradiance~\cite{brito_superradiance_2020}.
In the absence of a dissipating mechanism in the ergoregion, negative energy excitations grow, which leads to the splitting of the MQV into multiple moving SQVs~\cite{geelmuyden_sound-ring_2021,patrick_quantum_2022,patrick_origin_2022}. 

Typically, purely rotating geometries in conservative systems do not allow energy to dissipate inside the ergosurface (which bounds the ergoregion) and are thus intrinsically unstable~\cite{giacomelli_ergoregion_2020,patrick_origin_2022}.
Although MQVs in conservative systems may be stabilized by literally draining energy out---for example, by opening a funnel as in superfluid helium experiments~\cite{svancara_rotating_2024}---the resulting inward radial flow complexifies the mode structure in the ergoregion~\cite{visser_vortex_2005}, preventing investigation of the process at the origin of the dynamical instability of MQVs.

Here, we adopt another approach and generate stationary MQVs in purely rotating flows.
We use a driven dissipative quantum fluid of light in which the ergoregion instability rate is lower than the spatially homogeneous losses, preventing the complete development of the instability. This allows us to reach a stationary state in which the macroscopic phase is pinned by the driving field, while enabling the study of the instability precursor.

The spectrum of collective excitations of MQVs, which rules their quantum hydrodynamics, changes along the radius and is generically predicted to have negative energy trapped modes in the ergoregion, surrounded by a continuum of propagating positive energy modes at large radii~\cite{vieira_vortex_2025, giacomelli_ergoregion_2020, geelmuyden_sound-ring_2021, patrick_quantum_2022, patrick_origin_2022}.
Recent experiments in $C\sim 10^4$ MQVs of superfluid helium resolved this spectrum at large radii only~\cite{svancara_rotating_2024}, outside the ergoregion, and could not resolve it at a short radius.

Having stabilized the MQV, we adapt a recently developed coherent probe spectroscopy (CPS) method~\cite{claude_high-resolution_2022,claude_spectrum_2023} to measure the spatially resolved spectrum from large to short radii, including well inside the ergosurface all the way to the MQV core.
There we observe trapped negative energy excitations and thus fully characterize the generic perturbation spectrum for the first time.
This confirms the origin of MQV (in)stability in ergoregion physics and opens the way to the observation of rotational superradiance from ergosurfaces only.

\begin{figure}[ht]
    \centering
\includegraphics[width=\linewidth]{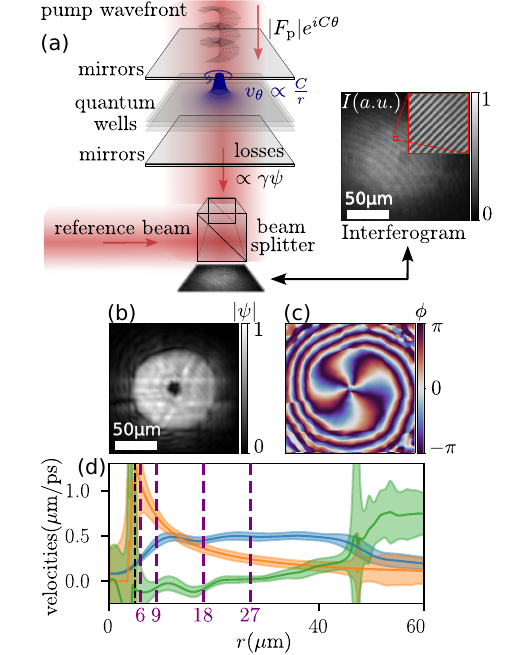}
    \caption{
    MQV generation and mean-field measurements.  
    (a) Schematic of the experimental setup. A continuous-wave laser with a helical wavefront pumps a microcavity near resonance, forming a MQV.
    A flat-phase reference beam, interferes with photonic signal exiting the cavity, allowing the full polariton field reconstruction by OAI \cite{SM}. (b), (c) Amplitude and phase of the polariton field.  
    (d) Azimuthally averaged velocity components $v_r$ (green), $v_\theta$ (orange) and $\sqrt{\hbar gn/m^{\star}}$ (blue).
    Shaded areas represent one standard deviation over the azimuthal angle.
    The black dashed line at $\SI{5}{\micro\meter}$ marks the approximate lower bound for the validity of the azimuthal average (inside that radius, the SQVs make the profile highly inhomogeneous azimuthally).
    Vertical dashed purple lines: radii at which the Bogoliubov spectrum is shown in Fig.~\ref{fig:w_m_disp}.}
    \label{fig:mean_field}
\end{figure}

\textit{Driven-dissipative quantum fluid of light---} Our quantum fluid is realized with microcavity exciton-polaritons (polaritons).
We coherently excite the cavity near resonance and obtain a 2D polariton fluid whose mean field $\psi(\bs{r}, t)$ is governed by the driven-dissipative Gross-Pitaevski equation~\cite{carusotto_quantum_2013}
\begin{multline}
    \ci\hbar\frac{\partial\psi(\bs{r},t)}{\partial t} = \ci\hbar F_\tp (\bs{r}, t)  -\ci\hbar\frac{\gamma}{2} \psi(\bs{r},t)\\
    + \left(\hbar\omega_0 -\frac{\hbar^2\nabla^2}{2m^\star}+ \hbar \left[gn(\bs{r},t) + g_\trr n_\trr(\bs{r},t)\right]\right)\psi(\bs{r},t),
    \label{eq:GPEpol}
 \end{multline}    
with $F_\tp (\bs{r}, t) = |F_\tp(\bs{r})|\ee^{\ci\left(\phi_\tp(\bs{r})-\omega_\tp t\right)}$ the electromagnetic field of the laser pump and $\gamma$ the decay rate of polaritons (that yield photons exiting the cavity at about the same rate).
$\omega_0$ is the frequency of polaritons at wavevector $\bs{k}=0$, $\hbar^2\nabla^2\psi/2m^\star$ their kinetic energy in the cavity plane, and $m^\star$ their effective mass.
$\hbar gn$ is the repulsive interaction energy, while $g_\trr n_\trr=\beta\times gn$ ($\beta=cst\geq0$) phenomenologically accounts for possible modifications to the interaction energy under the effect of a long-lived exciton reservoir not coupled to the pump field~\cite{amelio_galilean_2020}~\footnote{We have $\partial_t n_\trr = -\gamma_\trr n_\trr + \gamma_\mathrm{in} n$. 
where $\gamma_\trr$ is the long-lived exciton reservoir decay rate and $\gamma_\mathrm{in}$ the decay rate of polaritons into the reservoir. Taking the steady state of the reservoir rate equation shows that $n_\trr$ and $n$ are proportional through $\gamma_\trr n_\trr =  \gamma_\mathrm{in}n$.
This energy renormalization modifies the spectrum of collective excitations as well.
}.
Eq.~\eqref{eq:GPEpol} describes a driven-dissipative quantum fluid $\psi = |\psi|e^{i\phi}$ whose velocity depends on the field phase gradient by $\bs{v}=\hbar\boldsymbol{\nabla}\phi/m^{\star}$ and of density given by $n=|\psi|^2$ \cite{carusotto_quantum_2013}.

Optical control by coherent excitation allows us to engineer MQVs by shaping the pump field into an optical vortex $\phi_\tp(\bs{r})=C\theta$, forming stationary solutions of Eq.~\eqref{eq:GPEpol} of the form $\psi_C(\bs{r},t)=\sqrt{n(\boldsymbol{r})}\ee^{\ci(C\theta+S(r)-\omega_\tp t)}$ [see~\cite{SM} for remarks on spatial inhomogeneity] with azimuthal velocity profile
\begin{equation}
    \label{eq:velocityfield}
        v_\theta=\frac{\hbar}{m^\star}\frac{C}{r}.
\end{equation}

\textit{Mean field of a stationary $C=4$ MQV---}
In the experiment, we excite the cavity near resonance with a linearly polarized continuous-wave pump laser.
We control $\phi_\tp(\boldsymbol{r})$ with a spatial light modulator (SLM).

We measure the MQVs properties by performing off-axis interferometry (OAI) on light exiting the cavity at $\omega_\tp$, retrieving both the intensity and phase information of the mean field $\psi_C(\bs{r})$ [see~\cite{SM} and Fig.~\ref{fig:mean_field}~(a)].

Fig.~\ref{fig:mean_field}~(b) and (c) show the amplitude and phase of the mean field, which is stationary, while (d) shows $\sqrt{\hbar gn /m^\star}$ in blue \footnote{We use $\sqrt{\hbar gn / m^{\star}}$ as a proxy for the amplitude of the polariton field. This quantity has the dimension of a velocity and corresponds to the Bogoliubov hydrodynamic speed of sound in the case $\delta(v) = gn + g_\mathrm{res}n_\mathrm{res}$, where the Bogoliubov dispersion is linear at low $k$. However, as discussed in~\cite{falque_polariton_2024} and~\cite{SM}, the Bogoliubov field becomes massive when $gn + g_\mathrm{res}n_\mathrm{res} > \delta$, which is the case here. In this regime, the notion of a hydrodynamic speed of sound no longer makes sense.} and the azimuthally averaged velocities components $v_\theta$ in orange and $v_r=\frac{\hbar}{m^\star}\partial_rS$ in green (shaded areas, one standard deviation).
We distinguish three characteristic regions, visible in (b) and (d): the vortex core, with low amplitude; the vortex bulk, at intermediate radii with high amplitude; and the outer region ($r\gtrsim\SI{47}{\micro\meter}$), where the amplitude drops.
In the core, the azimuthal averaging in Fig~\ref{fig:mean_field}~(d) is not meaningful due to the strong spatial modulation induced by the position of the SQVs.
In the bulk, the phase exhibits the expected $4 \times 2\pi$ winding with a negligible radial gradient.
The radial velocity \(v_r\) remains null, while the azimuthal velocity follows the expected \(v_\theta \propto C/r\) profile.
At even larger radii, where the field amplitude drops as the pump amplitude vanishes, the polariton field is rapidly expelled, as evidenced by the steep radial phase gradient and the resulting high \(v_r\).
This behavior reflects the conversion of interaction energy into kinetic energy, in agreement with the quasi-conservation of the flow current~\cite{amelio_perspectives_2020}. 

\textit{Collective excitations of the vortex fluid---}
The quantum hydrodynamics of MQVs are ruled by the spectrum of low energy collective excitations, which we study with the Bogoliubov approach~\cite{bogolyubov_theory_1947}: we linearize Eq~\eqref{eq:GPEpol} around the stationary state $\psi_C$ and find the eigenmodes of linearized dynamics~\cite{giacomelli_ergoregion_2020,patrick_origin_2022,patrick_quantum_2022}. Defining the perturbed field by $\psi = \ee^{\ci\left(C\theta+S(r)-\omega_\tp t\right)}\left(f(r) + \delta\psi(\boldsymbol{r}, t)\, e^{-\gamma t/2} \right)$, the eigenmodes for the Bogoliubov field are azimuthal plane waves
\begin{equation}
\delta\psi(\bs{r}, t)=u(r)\ee^{\ci \left(m\theta-\omega_\tB t\right)}+v^*(r)\ee^{-\ci \left(m\theta-\omega_\tB t\right)}
\end{equation}
where the radial part $\ket{\delta\psi}:=\left( u(r), v(r)\right)^{T}$
is given by the Bogoliubov-de Gennes equation $(\hbar\omega_\tB-\hat{\mathcal{L}}_{C,m})\ket{\delta\psi}=0$~\cite{SM}. 

Previous theoretical work in conservative systems~\cite{patrick_quantum_2022} found that the exact solutions $\delta\psi(\bs{r}, t)$ are well approximated by the analytic Wentzel-Kramers-Brillouin (WKB) method~\cite{berry_semiclassical_1972}, that assumes a slowly varying mean-field on top of which Bogoliubov excitations behave locally as plane waves $\delta\psi_\mathrm{WKB}=A(r)\ee^{\ci\left( m\theta + \int p(r)dr-\omega_\tB^\pm t\right)}+B(r)\ee^{-\ci\left( m\theta + \int p(r)dr-\omega_\tB^\pm t\right)}$ with a local wavevector $\bs{k}=p(r)\hat{\bs{e}}_r+\frac{m}{r}\hat{\bs{e}}_\theta$ ($p$ is a radial wavevector) and a fixed angular frequency that satisfies the dispersion relation 
\begin{multline}
      \label{eq:lfdisp}
      \omega_\tB^\pm\Big(p, \frac{m}{r}\Big)=\frac{\hbar}{m^{\star}}  \frac{Cm}{r^2} + v_rp \\ \pm \sqrt{\left(\hbar\frac{m^2/r^2+p^2}{2m^\star}  -\delta(v) +2gn +g_\trr n_\trr  \right)^2-(gn)^2}
\end{multline}
in the laboratory frame, with $\frac{\hbar}{m^\star}\frac{Cm}{r^2}$ and $v_rp$ the azimuthal and radial Doppler shifts and $\hbar\frac{m^2/r^2+p^2}{2m^\star}$ the standard kinetic energy term~\footnote{Note that here we redefine $\delta$ from $\delta(\boldsymbol{k}_\tp) = \omega_\tp - \omega_0 - \hbar k_\tp^2 / 2m^{\star}$ to $\delta(v) = \omega_\tp - \omega_0 - m^{\star}v^2 / 2\hbar$. In the plane wave pumping case, due to strict phase locking, $\boldsymbol{v} \propto \boldsymbol{k}_\tp$, so this change is purely conventional. However, in the general case where the polariton phase is not strictly determined by the pump phase, the effective detuning should depend on $v$ rather than $\nabla\phi_\tp$~\cite{SM}.}.
Owing to the symmetry of the Bogoliubov problem~\cite{castin_bose-einstein_2001}, the spectrum exhibits the doubling $\omega_\mathrm{B}^+(p, m/r) = -\omega_\mathrm{B}^-(-p, -m/r)$, with both frequencies corresponding to the same excitation. Consequently, the dynamics can be fully captured by a single frequency branch; in the following, we retain $\omega_\mathrm{B} = \omega_\mathrm{B}^+$. Such an excitation on top of the mean field will be associated to a modification of the total energy given by $\delta E=Q\times\hbar\omega_\tB$, where $Q = \int r\,\di r(\abs{u}^2-\abs{v}^2)$ is a conserved quantity that is positive for the frequency solutions we consider~\cite{bogolyubov_theory_1947,castin_bose-einstein_2001}.
The sign of this energy, carried by the perturbation, is then given by the sign of $\hbar\omega_\tB$.

In the following, we evaluate the agreement between the WKB approximation and the measured spectrum in an inhomogeneous system.
In particular, we expect that the variations of the mean field quantities along the radius (notably in the core) will cause a breakdown of the WKB solutions and the occurrence of turning points in Hamilton's equation for the excitations~\cite{patrick_rotational_2021}.
Eq.~\eqref{eq:lfdisp} does not account for these modified boundary conditions and, for example, the discretization of the spectrum in $p$ that we will observe later.

\begin{figure}[t]
    \centering
\includegraphics[width=\linewidth]{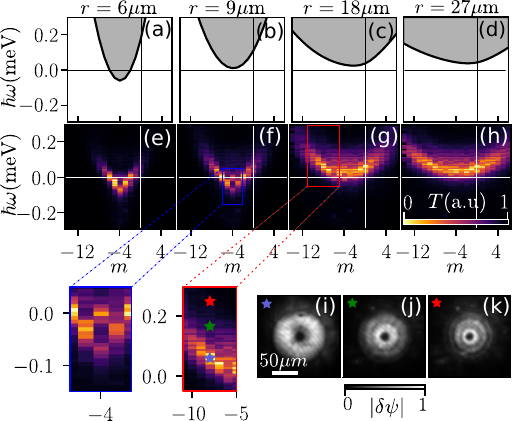}
    \caption{Spatially resolved Bogoliubov spectrum.
    (a)- (d) WKB spectrum $\hbar\omega_\mathrm{B}$ at $r = 6$, $9$, $18$, and $\SI{27}{\micro\meter}$. Solid line, $p=0$ mode; shaded area, $|p|>0$ modes.
    (e) - (h) $(\hbar\omega, m)$ probe transmission intensity normalized radius by radius at $r = 6$, $9$, $18$, and $\SI{27}{\micro\meter}$, plotted in dashed purple lines in Fig. \ref{fig:mean_field} (d).
    For a given $r$, each pixel value is obtained by summing the squared probe amplitude over the interval $[r - \SI{3}{\micro\meter}, r + \SI{3}{\micro\meter}]$. (i)-(k) Normalized probe amplitude for $m=-8$ at three different $\hbar\omega$ with correspondingly color-coded star labels in the inset of (g).
    }
    \label{fig:w_m_disp}
\end{figure}

Fig.~\ref{fig:w_m_disp}~(a)-(d) shows the WKB spectrum~\eqref{eq:lfdisp} $\hbar\omega_\tB(p,\frac{m}{r})$ at different radii for $p=0$ (black line) and $\vert p\vert>0$ (shaded area).

At high $m$, the dispersion is dominated by the quadratic term $\hbar\frac{m^2/r^2}{2m^\star}$, leading to a parabolic shape at high $m$ for all radii.
At lower $m$, the Doppler effect becomes significant.
While the radial Doppler term $v_rp$ is negligible in the core and the bulk of the vortex, the azimuthal contribution is linear in $m$ with a positive slope given by $\frac{\hbar}{m^{\star}}\frac{C}{r^2}$, tilting the dispersion counter clockwise and breaking the $m \leftrightarrow -m$ symmetry.
At high radii, $\SI{27}{\micro\meter}$ and $\SI{18}{\micro\meter}$ in Fig.~\ref{fig:w_m_disp}~(d) and~(c), this slope is rather low, so the spectrum is entirely at positive frequency.
Nearer to the vortex core (at $\SI{6}{\micro\meter}$, Fig.~\ref{fig:w_m_disp}~(a)) the Doppler term tilts the dispersion enough for part of the spectrum to be dragged to negative frequencies.
The surface that separates the region where the spectrum is only at positive frequencies (positive energy modes only) from the region where part of the spectrum is at negative frequencies is called the ergosurface.
Its radius $r_e$ is where $\min_{p,m}\Big\{\hbar\omega_\tB\Big(p, \frac{m}{r}\Big)\Big\}=0$.
In the ergoregion, negative energy modes can be excited.

\begin{figure*}[ht!]
    \centering
\includegraphics[width=\linewidth]{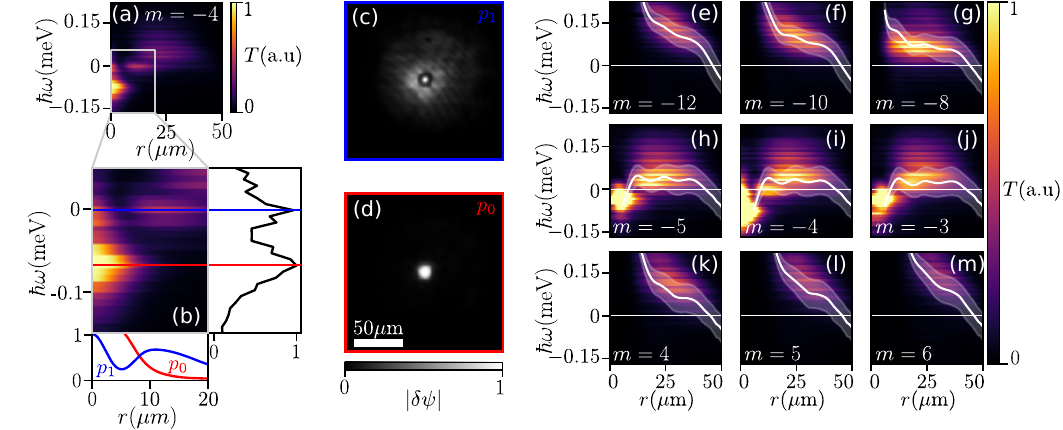}
    \caption{
    Effective potential model.  
    (a) $(\hbar\omega$--$r)$ probe transmission intensity for $m = -4$.
    (b) Focus on the core modes. Right, integrated transmission for $r < 20~\mu\mathrm{m}$; bottom, azimuthally averaged probe intensity of the $p_0$ (red) and $p_1$ (blue) modes.
    Note that the $p_0$ mode saturates the detector for $r < \SI{6}{\micro\meter}$; these data points exceed 1 in arbitrary units and are omitted.  
    (c) - (d) Rescaled density of the $p_1$ and $p_0$ modes for $m=-4$.
    (e)- (m) $(\hbar\omega$--$r)$ probe transmission intensity for different $m\in[-12,6]$ (common color scale). White line, azimuthal average of $\mathrm{min}\{\hbar\omega_B^+\}$, corresponding to $V_m(r)$~\eqref{eq:eff_pot}; shaded area, one standard deviation.
    Note that panels (h) - (j) are saturated under the chosen color scale to capture all features of interest.
    } 
    \label{fig:trapped_modes}
\end{figure*}

\textit{Observation of the Bogoliubov spectrum ----}
CPS is implemented with another cross-polarized continuous-wave probe laser of intensity two orders of magnitude below $\abs{F_\tp(\bs{r})}^2$.
The probe spatial mode is also a vortex beam that overlaps the entire pump mode.
Its phase $\phi_\tpr(\bs{r})= \phi_\tp(\boldsymbol{r})+ m\theta$ is controlled with an SLM.
For each \(m\), we scan the probe frequency $\omega_{\mathrm{pr}}$, and Bogoliubov excitations are resonantly generated at radius \(r\) when the excitation energy $\hbar\omega \equiv \hbar(\omega_{\mathrm{pr}} - \omega_{\mathrm{p}}) = \hbar\omega_\tB(p,\frac{m}{r})$.
At this point, the probe light is transmitted through the microcavity. This transmitted probe light being proportional to the Bogoliubov field, by OAI at $\omega_\tpr$ we retrieve only the $+\omega_\tB$ component $u(r)\ee^{\ci(m\theta-\omega_\tB t)}$ of $\delta\psi$~\footnote{We observe the $\omega_\tB$ component of the mode, i.e. $u(r)\ee^{\ci\left(m\theta-\omega_\tB t\right)}$ because OAI can only resolve optical fields of the same frequency as the probe. We cannot observe the four-wave mixing signal $v(r)\ee^{-\ci\left(m\theta-\omega_\tB t\right)}$ as in~\cite{falque_polariton_2024} because it exits the cavity at $2\omega_\tp-\omega_\tpr$ and therefore does not interfere with the reference beam at $\omega_\tpr$.}.
We then obtain the radially resolved spectrum by means of radial masks applied to the fields data upon processing. 

Figure~\ref{fig:w_m_disp}~(e) - (h) shows the spectrum of Bogoliubov excitations.
At \(r = \SI{27}{\micro\meter}\) and \(r = \SI{18}{\micro\meter}\), the dispersion lies entirely at positive frequencies for all \(m\) modes. For each \(m\), the dominant feature is a broad \(p = 0\) resonance whose center defines its frequency and whose width is associated to its decay rate. Higher-order \(|p| > 0\) excitations are also present at higher frequencies, forming a weak and diffuse halo above the primary modes, as can be seen in the inset of Fig.~\ref{fig:w_m_disp}~(g).
The amplitudes for m=-8 for the primary mode (blue star) together with two higher frequency modes of the halo (green and red stars) are shown in Fig.~\ref{fig:w_m_disp}~(i)-(k).
The primary mode density is flat in the bulk, corresponding to p=0, while the amplitude of higher frequency modes is modulated along the radius, corresponding to higher $|p|>0$ modes.

In the bulk, the agreement between the WKB approximation and the measured spectrum is qualitatively good.
As anticipated, closer to the core (where the mean field inhomogeneity is large) we observe differences between the two.
In particular, at \(r = \SI{9}{\micro\meter}\), although the WKB approximation predicts positive energy modes, we observe that $\hbar\omega_\tB<0$ from \(m = -5\) to $-3$, showing that these $m$ modes are negative energy modes, a clear signature of the formation of the ergoregion. For smaller radii, the negative energy modes are also present, as expected.

The observation of negative energy modes in the core of the system proves that we have generated a stationary MQV in which the ergoregion instability associated with the excitation of these negative energy modes is quenched by the finite polariton lifetime.

\textit{Effective potential and core modes---} The structure of the spectrum of collective excitations changes with $r$.
As we can see in Fig.~\ref{fig:w_m_disp}~(g) and (h), the spectrum is continuous for all $m$ outside the ergosurface.
In contrast, in the ergoregion the spectrum is continuous for all azimuthal numbers except $m = -4$.
The inset of Fig.~\ref{fig:w_m_disp}~(f) shows that two discrete resonances emerge at $\hbar\omega \approx 0\mathrm{meV}$ and $\hbar\omega \approx-0.07\mathrm{meV}$.

Figure~\ref{fig:trapped_modes}~(a) shows the probe transmission intensity for $m=-4$ as a function of $\hbar\omega$ and $r$.
In the core, we observe the same two resonances, visible in the gray rectangle box.
These correspond to two different radial modes, labeled \(p_0\) and \(p_1\) for the lowest- and highest-frequency mode, respectively.
They are isolated from the rest of the otherwise continuous spectrum, which lies at a larger radius and higher frequency.

Numerical studies in conservative quantum fluids~\cite{patrick_origin_2022,patrick_quantum_2022} used the WKB framework to demonstrate that these resonances are radially trapped modes in the effective potential
\begin{equation}
\label{eq:eff_pot}
V_m(r) = \min_p\hbar\omega_\tB\left(\frac{m}{r}, p\right).
\end{equation}
In our case, since the radial Doppler contribution is negligible, we obtain the minimum in $p$ for $p=0$, so $V_m(r)=\frac{\hbar}{m^\star}\frac{Cm}{r^2}+\sqrt{\left(\hbar \frac{m^2/r^2}{2m^\star}-\delta(v)+2gn +g_\mathrm{res}n_\mathrm{res} \right)^2-(gn)^2}$.
For a given \(m\) and $\hbar\omega$, propagation is allowed only in the regions where \(\hbar\omega > V_m(r)\).
Whenever \(\hbar\omega \leq V_m(r)\), $p$ becomes imaginary and the modes evanescent, so that the radius $r^\star$ (where \(\hbar\omega = V_m(r^\star)\)) acts as a reflective boundary for propagating waves.
For example, for \(m = -4\), the propagation of $\hbar\omega\leq 0$ modes is restricted to \(r \lesssim \SI{10}{\micro\meter}\).
Similarly to cavity modes~\cite{patrick_origin_2022}, only a discrete set of radial standing waves indexed by a radial wavenumber \(p\), each at a well-defined \(\hbar{\omega_\mathrm{B}}_p\), can oscillate in that region.

Fig.~\ref{fig:trapped_modes} (b) zooms in on the low frequency and short radius part of Fig.~\ref{fig:trapped_modes} (a) and shows the azimuthally averaged amplitude of the two modes localized in the core, \(p_0\) in red and \(p_1\) in blue.
For $p_1$, we observe a bright resonance near $r=0$ and a dimmer one at $r=\SI{11}{\micro\meter}$ (two-peak blue profile below (b)), indicating of a standing wave with two antinodes.
This is also visible in Fig.~\ref{fig:trapped_modes} (c), which shows the $p_1$ mode density in the microcavity plane. On the other hand, $p_0$ exhibits only one antinode at $r=0$ (single peak red profile below (b)), as we see in Fig.~\ref{fig:trapped_modes}~(d).
The spatial structure and frequency discretization of the $m=-4$ spectrum are clear signatures of trapped modes in an effective cavity inside the core.

Fig.~\ref{fig:trapped_modes} (e)-(m) shows the probe transmission intensity along $r$ and $\hbar\omega$ for various $m$.
The effective potential~\eqref{eq:eff_pot} is shown in white.
We see that for $m>0$ and for $m\leq-8$, the potential increases monotonically toward the center, preventing radial trapping. As propagation is allowed only for \(r > r^\star\), no probe is transmitted near the core.

For \(m > 0\), Eq.~\eqref{eq:eff_pot} states that both the Doppler shift and the quadratic azimuthal term $\hbar\frac{m^2/r^2}{2m^\star}$ increase \(V_m(r)\) as $r\rightarrow 0$, reinforcing the monotonic profile.
For \(m < 0\), these contributions compete: the Doppler term is negative, so it shifts the dispersion downward, whereas the quadratic azimuthal term pushes it upward.
At low \(|m|\), the Doppler term dominates and allows a cavity to form; at higher \(|m|\) ($m<=-8$), the quadratic azimuthal term prevails, suppressing the trapping region.

The range of $m$ values for which cavity modes can form, as well as the number of modes supported within the cavity, is then governed by the shape of the effective potential~\eqref{eq:eff_pot}, which ultimately depends on the excitation parameters \(C\), $\delta(v(r))$ and $\abs{F_\tp(r)}^2$.
Remarkably, as predicted for infinite size conservative systems~\cite{giacomelli_ergoregion_2020}, the deepest effective cavity that produces the optimal trapping condition is obtained for \(m = -C=-4\).

\textit{Discussion---}
We observed generic features of rotating geometries, including negative energy modes inside the ergoregion and trapped modes deep within the vortex core.
The qualitative correspondence we observed between the probe response and WKB calculations~\cite{giacomelli_ergoregion_2020,patrick_origin_2022,patrick_quantum_2022} confirms theoretical predictions based on this framework.
It also confirms the origin of the intrinsic MQV instability in purely rotating geometries in ergoregion physics: namely, the trapping of negative energy waves (whose amplitude can grow exponentially if not dissipated) therein.
Alternatively to the spectral measurements presented here, optical control over the probe field could enable the study of other universal features in geometries creating an effective potential, such as light ring oscillations~\cite{torres_quasinormal_2020,svancara_rotating_2024} or rotational superradiance~\cite{torres_rotational_2017,braidotti_measurement_2022}.
Quantum optics methods could, in turn, enable the observation of the associated generation of entanglement~\cite{delhom_entanglement_2024}.

In our driven-dissipative quantum fluid of light, additional knobs are available that allow for broad exploration and control beyond what could be realized in conservative systems traditionally studied.
For example, we created stable MQVs without a central drain --a highly inhomogeneous system-- and spatially resolved the spectrum of collective excitations.
Our spectroscopic method~\cite{claude_high-resolution_2022} complemented by off-axis interferometry, which allowed us to go beyond previous measurements~\cite{stepanovDispersionRelationCollective2019,pieczarka_observation_2020,estrecho_low-energy_2021,falque_polariton_2024}, could be used further to investigate the quantum hydrodynamics on other inhomogeneous geometries where analytical descriptions are not available.

The data and analysis scripts that support the findings of this article are openly available in~\cite{zenodo}.

\begin{acknowledgements}
    We thank Luca Giacomelli, Ivan Agullo, Iacopo Carusotto, Adria Delhom and Myrann Baker Rasooli for sharing their insights on fluids of light and the physics of quantized vortices therein. We are particularly grateful to Adria Delhom for his careful reading of the manuscript. We acknowledge funding from the EU Pathfinder 101115575 Q-One and from CNRS via a 80prime PhD studentship. AB acknowledges support from the Institut Universitaire de France.
\end{acknowledgements}

\end{document}

\title{Supplemental Material for: Multiply quantized vortex spectroscopy in a quantum fluid of light}

\author{Killian Guerrero}\affiliation{Laboratoire Kastler Brossel, Sorbonne Universit\'{e}, CNRS, ENS-Universit\'{e} PSL, Coll\`{e}ge de France, Paris 75005, France}
\author{K\'evin Falque}\affiliation{Laboratoire Kastler Brossel, Sorbonne Universit\'{e}, CNRS, ENS-Universit\'{e} PSL, Coll\`{e}ge de France, Paris 75005, France}
\author{Elisabeth Giacobino}\affiliation{Laboratoire Kastler Brossel, Sorbonne Universit\'{e}, CNRS, ENS-Universit\'{e} PSL, Coll\`{e}ge de France, Paris 75005, France}
\author{Alberto Bramati}\email{alberto.bramati@lkb.upmc.fr}\affiliation{Laboratoire Kastler Brossel, Sorbonne Universit\'{e}, CNRS, ENS-Universit\'{e} PSL, Coll\`{e}ge de France, Paris 75005, France}
\author{Maxime J. Jacquet}\email{maxime.jacquet@lkb.upmc.fr}\affiliation{Laboratoire Kastler Brossel, Sorbonne Universit\'{e}, CNRS, ENS-Universit\'{e} PSL, Coll\`{e}ge de France, Paris 75005, France}
\maketitle

\section{Experimental implementation}\label{app:setup}

\begin{figure}[!ht]
    \centering
\includegraphics[width=0.8\textwidth]{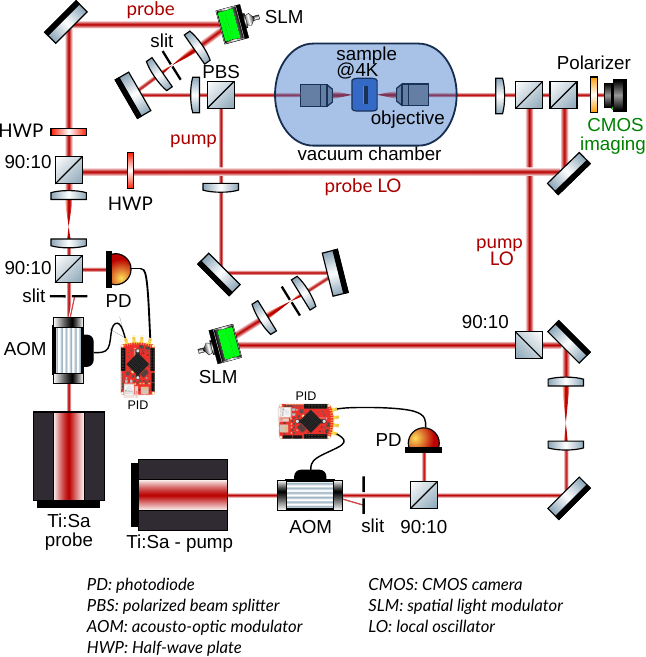}
    \caption{Experimental Setup}
    \label{fig:setup}
\end{figure}

\subsection{All-optical control and measurement of the mean field}
The polariton fluid is generated using a continuous wave (CW) Ti:Sapphire laser with a sub-MHz linewidth (Spectra-Physics Matisse), tuned to near-resonance with a semiconductor microcavity. The linearly polarized Gaussian beam emitted by the laser is first shaped using a phase-only LCOS spatial light modulator (SLM). We use the SLM to transform the Gaussian beam into a vortex beam by imprinting a target phase profile $\phi_{\mathrm{target}} = C\theta$, where $\theta$ is the azimuthal angle in a polar coordinate system centered on the SLM. To achieve full phase modulation, the target phase is combined with a superimposed vertical blazed grating. The first diffraction order is then selected in Fourier space using a pinhole placed after the SLM. The modulated beam is sent to the microcavity via a 4f telescope followed by a lens and a microscope objective $\times 50$, resulting in a total demagnification factor of 40.
In the plane of the microcavity, the pump beam is a vortex beam of waist $\SI{75}{\micro\meter}$, featuring a hole around $r=0$ in the intensity profile and a helical phase structure carrying a $C \times 2\pi$ phase winding.

The semiconductor microcavity consists of three InGaAs quantum wells embedded in a single-mode optical cavity formed by two highly reflective planar GaAs/AlGaAs Bragg mirrors with a finesse of 3000. The sample is engineered so that the quantum wells, separated by GaAs barriers, are positioned at the three antinodes of the cavity field to maximize light–matter interaction. The entire structure is placed in a closed-cycle liquid helium cryostat and cooled to temperatures $\lesssim\SI{10}{\kelvin}$. Under these conditions, cavity photons strong couple with quantum well excitons and the system eigenstates become two mixed modes of hybrid photon-exciton quasiparticules, the polaritons.

The angular momentum of the optical field is transferred to the polariton field, causing the system to behave like a rotating fluid.
Polaritons have a lifetime of approximately $\tau = \SI{10}{\pico\second}$ in the cavity, dominated by photonic radiative losses.
The photonic field (losses) is proportional to the intra-cavity polariton field, meaning that both the amplitude and the velocity information are encoded in the losses. 
This photonic field is collected using a symmetric imaging setup (a $\times 50$ objective followed by a lens and a 4f telescope) and detected on a CMOS camera with an exposure time $t_\mathrm{exp}=\SI{100}{\micro\second}$. This setup allows us to record and average over $t_\mathrm{exp}/\tau\approx10,000$ realizations of the system dynamics -- we only see averaged dynamics.

To access both the phase and the amplitude of the field, we use an off-axis interferometry technique (OAI). This enables simultaneous reconstruction of the phase and the intensity of the photonic field. Although phase information directly gives the velocity of the fluid, extracting the density from the photonic intensity is more complicated, and a calibration must be performed using spectral measurements, as detailed in Appendix~\ref{app:soundspeed}.
In the end, the full field is resolved.

\subsection{All-optical control and measurement of the Bogoliubov field}
We adapted the coherent probe spectroscopy method introduced in~\cite{claude_high-resolution_2022} to characterize the Bogoliubov excitations of our inhomogeneous fluid.

The Bogoliubov excitations are probed with another CW Ti:Sapph laser (M2 Solstis) whose frequency can be scanned continuously over a range of 150MHz. The probe beam is crossed-polarized with the pump beam and has the same waist in the cavity plane, while its intensity is 2 to 3 orders of magnitude below.

We used another phase-only SLM to shape the probe beam with a $m\theta$ phase and the same optics as for the pump to image it onto the cavity.

The detection path is common with the mean-field detection scheme and weak probe photonic signals (losses) are measured again with the OAI method, with a reference obtained from a pick-up of the probe beam before the SLM.
We can thus monitor the probe transmission and identify the resonances giving the Bogoliubov spectrum.
Again, we have access to the full field.

\section{Polariton Fluid under plane wave excitation}

The lower polariton dynamics coupled to a dark excitonic reservoir is given by the following coupled equations \cite{stepanovDispersionRelationCollective2019,amelio_galilean_2020,claude_spectrum_2023}
\begin{align}\label{eq:dGPEannex}
\ci{{\hbar}}\partial_t\psi&=\lrsq{{{\hbar}}\omega_0-\frac{\hbar^{{2}}}{2m^\star}\nabla^2+\hbar g|\psi|^2+\hbar g_\mathrm{res}n_\mathrm{res}-\ci\frac{{{\hbar}}(\gamma_{\mathrm{LP}}+\gamma_\mathrm{in})}{2}}\psi+ \ci\hbar F_\tp\\
\partial_t n_\mathrm{res} &= -\gamma_\mathrm{res} n_\mathrm{res} + \gamma_\mathrm{in} n,
\end{align}
where $\psi$ is the lower polariton field function, $\omega_0$ the $k=0$ resonance, $m^{\star}$ the effective mass of the system and $F_\mathrm{p}$ is the driving laser field inside the cavity. The field is coupled incoherently with a dark reservoir via a loss channel. This reservoir contributes to the interactions via the $\hbar g_{\mathrm{res}} n_{\mathrm{res}}$ term. These equations can be written in terms of density phase variables $\psi = \sqrt{n}e^{i\theta}$, leading to a continuity equation and a Euler equation for a non-barotropic fluid \cite{carusotto_quantum_2013}\cite{falque_polariton_2024} whose velocity and density are given by $\boldsymbol{v}=\frac{\hbar}{m^{\star}}\boldsymbol{\nabla}\theta$ and $n$, respectively. 

Under plane wave excitation $F_\tp = |F_\tp|e^{i(\boldsymbol{k}_\tp \cdot \boldsymbol{r} - \omega_\tp t)}$ the polariton field will take the form $\psi = \sqrt{n}e^{\ci\left(\boldsymbol{k}_\mathrm{p}\cdot\boldsymbol{r}-\omega_{\mathrm{p}} t\right)}$ with
\begin{equation}
 n \left[ \left(\delta(v) +  g_{\mathrm{res}} n_{\mathrm{res}} +  g n \right)^2 + \left( \frac{\gamma}{2} \right)^2 \right] = \hbar^2|F_\tp|^2,
 \label{eq:bistab}
\end{equation}
where we defined $\boldsymbol{v} = \frac{\hbar}{m^{\star}}\boldsymbol{k}_{\mathrm{p}}$, $\delta = \omega_\tp - \omega_0 -  m^\star v^2/2\hbar$ and $\gamma = \gamma_\mathrm{LP}+\gamma_\mathrm{in}$. Here, due to the conservation of the wave vector, the velocity of the fluid is proportional to $\boldsymbol{k}_\tp$, so $\delta$ ultimately depends on $\boldsymbol{k}_\tp$. However, in the case of arbitrary coherent excitation, with a spatially inhomogenous phase, phase locking is no longer strict and the detuning becomes a function of the local flow velocity $\boldsymbol{v}=\frac{\hbar}{m^{\star}}\boldsymbol{\nabla}\phi$. Hence, we write $\delta \equiv \delta(v)$ with $\delta(v) = \omega_\tp - \omega_0 - \frac{m^\star v^2}{2\hbar}$.

Eq.~\ref{eq:bistab} gives the nonlinear relation between $|F_\tp|^2$ and $n$.
We work in the regime where $\delta/\gamma>\sqrt{3}/2$ so we have a bistability cycle~\cite{baas_optical_2004} with high- and low-density states, see Fig.\ref{fig:SM_simu}.

In the case of plane-wave excitation, linear perturbations around $\psi = \sqrt{n}\ee^{\ci\left(\boldsymbol{k}_\tp\cdot\boldsymbol{r}-\omega_\tp t\right)}$, defined by $\psi_{tot} = \ee^{\ci\left(\boldsymbol{k}_\tp\cdot\boldsymbol{r}-\omega_\tp t\right)}(\sqrt{n}+\delta\psi\ee^{-\gamma t/2})$ can be decomposed in plane wave eigenmodes whose dispersion relation is given by
\begin{equation}
      \label{eq:lfdisp}
      \omega_\mathrm{B}^\pm(\boldsymbol{k})=\boldsymbol{v}\cdot\boldsymbol{k}\pm\sqrt{\left(\frac{\hbar k^2}{2m^\star} -\delta(v) + 2gn +g_\mathrm{res}n_\mathrm{res}  \right)^2-(gn)^2}.
\end{equation}

This is the well-know Bogoliubov dispersion of a plane-wave polariton field. However, when the pump deviate from a plane-wave and thus the mean-field acquires an inhomogeneous density or velocity, this global Bogoliubov dispersion~\eqref{eq:lfdisp} no longer applies.

\section{Cross-polarization detection scheme}

In our experiment, the Bogoliubov field is created with a probe of intensity 2 to 3 orders of magnitude below the pump one. To ensure a good detection of this Bogoliubov field without saturating the CMOS camera, we set the probe crossed-polarized configuration so that can better separate it. In that case, the polarization degree of freedom of the polariton need to be included in the model since different polarization species will interact differently However, we show here that we can reduce the problem to a scalar description. 

We work in the circular polarization basis $(\sigma^+, \sigma^-)$. In this basis, any linearly polarized field can be written as a superposition of equal-amplitude $\sigma^+$ and $\sigma^-$ components with a relative phase. A horizontally polarized field corresponds to $E_H = \frac{1}{\sqrt{2}}(E_+ + E_-)$, while a vertically polarized field corresponds to $E_V = \frac{1}{\sqrt{2}}(E_+ - E_-)$. This implies $E_+ = E_-$ for a horizontally polarized field and $E_+ = -E_-$ for a vertically polarized one. Therefore, in the circular basis, the pump satisfies $F_{\tp_+} = F_{\tp_-}$. This creates a polariton field composed of two components whose mean fields $\psi_{0\pm}$ are governed by:
\begin{align}
\ci\hbar\partial_t\psi_{0+} &=
\left[\,\hbar\omega_0 - \frac{\hbar^2}{2m^\star}\nabla^2 + \hbar g_{++}|\psi_{0+}|^2 + \hbar g_{+-}|\psi_{0-}|^2 + \hbar g_{\mathrm{res}} n_{\mathrm{res}} - \ci\frac{\hbar(\gamma_{\mathrm{LP}} + \gamma_{\mathrm{in}})}{2}\,\right] \psi_{0+}
+ \ci \hbar F_{\tp_+}, \\
\ci\hbar\partial_t\psi_{0-} &=
\left[\,\hbar\omega_0 - \frac{\hbar^2}{2m^\star}\nabla^2 + \hbar g_{--}|\psi_{0-}|^2 + \hbar g_{+-}|\psi_{0+}|^2 + \hbar g_{\mathrm{res}} n_{\mathrm{res}} - \ci\frac{\hbar(\gamma_{\mathrm{LP}} + \gamma_{\mathrm{in}})}{2}\,\right] \psi_{0-}
+ \ci \hbar F_{\tp_-}, \\
\partial_t n_{\mathrm{res}} &= -\gamma_{\mathrm{res}} n_{\mathrm{res}} + \gamma_{\mathrm{in}} \big(|\psi_{0+}|^2 + |\psi_{0-}|^2\big),
\end{align}
where $g_{++}$ is the interaction constant for polaritons of the $\sigma^+$ component, $g_{--}$ is the interaction constant for polaritons of the $\sigma^-$ component and $g_{+-}$ is the interaction constant for polaritons of different components. In our sample, $g_{++}=g_{--}\equiv g_\mathrm{intra}$ and $g_{+-} \ll g_\mathrm{intra}$. For simplicity, let us set $g_\mathrm{+-}=0$, which decouples the equations, although the detection scheme also holds for finite $g_{+-}$. In that case, the equations of motion become

\begin{align}
\ci\hbar\partial_t\psi_{0\pm} &=
\left[\,\hbar\omega_0 - \frac{\hbar^2}{2m^\star}\nabla^2 + \hbar g_{\mathrm{intra}}|\psi_{0\pm}|^2 + \hbar g_{\mathrm{res}} n_{\mathrm{res}} - \ci\frac{\hbar(\gamma_{\mathrm{LP}} + \gamma_{\mathrm{in}})}{2}\,\right] \psi_{0\pm}
+ \ci \hbar F_{\tp_\pm}, \\
\partial_t n_{\mathrm{res}} &= -\gamma_{\mathrm{res}} n_{\mathrm{res}} + \gamma_{\mathrm{in}} \big(|\psi_{0+}|^2 + |\psi_{0-}|^2\big).
\end{align}

Given that $F_{\tp_+} = F_{\tp_-} = F_{\tp}$, the equations are now identical and uncoupled for the two components. Therefore, the steady-state solutions satisfy
\begin{equation}
\psi_{0+} = \psi_{0-} \equiv \psi_0.
\end{equation}

Now let us consider a linearly vertically polarized probe on top of this mean field, meaning that $F_{\mathrm{pr}_+} = -F_{\mathrm{pr}_-}$ for the components in the $(\sigma^+, \sigma^-)$ basis. The $\sigma^+$ component $F_{\mathrm{pr}_+}$ will create a weak perturbation on top of the $\sigma^+$ component of the mean field and the $\sigma^-$ component $F_{\mathrm{pr}_-}$ will create a weak perturbation on top of the $\sigma^-$ component of the mean field such that
\begin{equation}
\psi_\pm = \psi_0  \pm \delta\psi.
\end{equation}

Since the detection is performed in the vertical polarization channel, proportional to $\psi_+ - \psi_- = 2\delta\psi$, the background $\psi_0$ is cancelled out and we directly measure the perturbation field.

Thus, in practice, we generate two uncoupled identical polariton fluids and probe them with equal and opposite probes in parallel. At detection, their background cancels due to polarization filtering, and only the perturbation appears in the signal. This provides clean access to the Bogoliubov field without contamination from the strong background fluid.

\section{Polariton field control under vortex excitation -- Numerical simulations}\label{app:numerics}

Unlike in the plane-wave case, where analytical steady-state solutions of Eq.~\eqref{eq:dGPEannex} exist, no analytical solution can be derived under finite-size optical vortex pumping $F_\tp = F_0(r) e^{-(r/w)^2} e^{\ci\left(C\theta-\omega_{\mathrm{p}} t\right)}$. In this case, Eq.~\eqref{eq:dGPEannex} must be solved numerically. To this end, we implement a spectral split-step method to simulate the system dynamics (Eq.\eqref{eq:dGPEannex} with $\gamma_{\mathrm{in}} = 0$).

\begin{figure}[ht!]
    \centering
\includegraphics[width=0.8\textwidth]{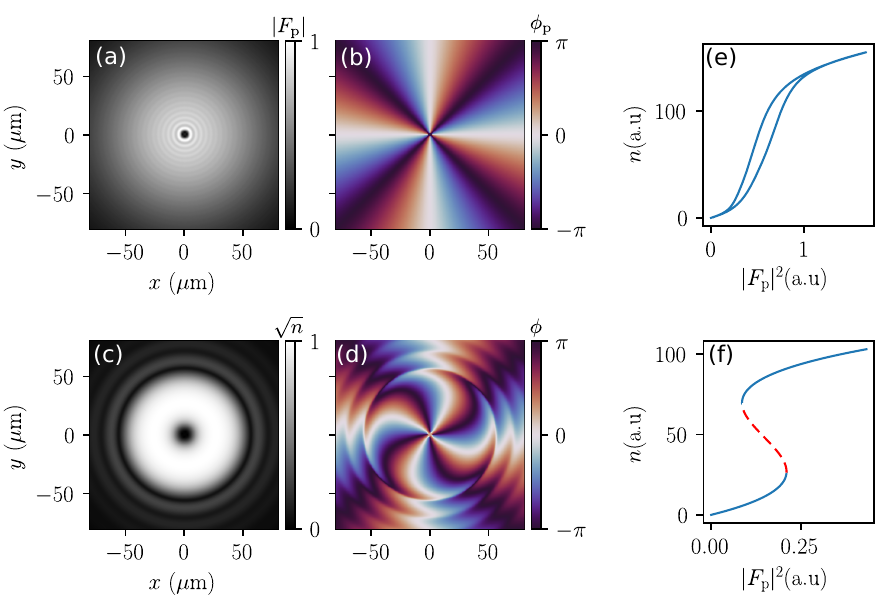}
    \caption{\textbf{Polariton field control with optical vortex pumping}.
    \textbf{(a)~(b)} Amplitude $|F_\mathrm{p}|$ and phase $\phi_\mathrm{p}$ of the optical vortex pump. \textbf{(c)~(d)} Amplitude $\sqrt{n}$ and phase $\phi$ of the steady-state polariton field obtained with an optical vortex pump.
    \textbf{(e)} Integrated density $(\int_{r<\SI{50}{\micro\meter}}n(r)\mathrm{d}r)$ as a function of the integrated pump intensity $(\int_{r<\SI{50}{\micro\meter}}\mathrm{d}r |F_\tp(r)|^2)$ when the overall amplitude $F_0$ is ramped up and down.
    \textbf{(f)} Homogeneous density $n$ as a function of the pump intensity $|F_\tp|^2$ in the plane wave excitation case, given by Eq.~\ref{eq:bistab}. The red dotted part is an unstable density state solution of Eq.~\ref{eq:bistab}
    }
\label{fig:SM_simu}
\end{figure}

Fig~\ref{fig:SM_simu}~(a),~(b) show the optical pump amplitude $|F_\tp|$ and phase $\phi_\tp$ maps, while (c)(d) show the resulting amplitude $\sqrt{n}$ and phase $\phi$ maps.
For $r<\SI{30}{\micro\meter}$, where $n$ is high and far from $r\approx\SI{50}{\micro\meter}$ where it drops, the polariton phase locks to the pump one, showing efficient transfer of the optical phase to the quantum fluid.
In contrast, in the vicinity of the density drop and in the outer region $r>\SI{50}{\micro\meter}$ where $n$ is low, the polariton phase develops a strong radial gradient. This indicates an outward flow of the field that can be understood in terms of quasi-conservation of the current~\cite{amelio_perspectives_2020}.

Fig.~\ref{fig:SM_simu}~(e) shows the integrated density $\int_{r<\SI{50}{\micro\meter}} n(\boldsymbol{r})\mathrm{d}\boldsymbol{r}$ as a function of the integrated pump intensity $\int_{r<\SI{50}{\micro\meter}}|F_\tp(\boldsymbol{r})|^2\mathrm{d}\boldsymbol{r}$ obtained by increasing and then decreasing the overall pump amplitude $F_0$. For the parameters we used here ($\delta(v=0)/\gamma>\sqrt{3}/2$), this curve is a hysteresis loop. Fig.~\ref{fig:SM_simu}~(f) shows that this is qualitatively the same result as in the plane wave case, Eq.~\ref{eq:bistab}.

Although neither strict phase locking nor analytical relations between $n$ and $|F_\tp|^2$ hold under finite-size vortex excitation, we observe the same qualitative behaviors. This indicates that control over the polariton field can still be achieved by tuning $|F_\tp|^2$ and $\delta(v=0)$ for $n$ and $\phi_\tp$ for $\phi$.

\section{Bogoliubov excitations on top of an inhomogeneous polariton flow} \label{app:bogotheo}

\subsection{General case}

The dynamics of lower polaritons coupled to a dark excitonic reservoir are described by the coupled equations \eqref{eq:dGPEannex}.
Under continuous-wave (CW) excitation $F_\tp = |F_\tp|e^{i(\phi_\tp(\boldsymbol{r}) - \omega_\tp t)}$, we seek stationary solutions of the form $\psi_0(r,t) = \sqrt{n_0(r)} e^{i(\phi_0(\boldsymbol{r}) - \omega_\tp t)}$. In this regime, the reservoir density satisfies $n_\mathrm{res} = \frac{\gamma_\mathrm{in}}{\gamma_\mathrm{res}} |\psi|^2$, and Eq.~\eqref{eq:dGPEannex} simplifies to
\begin{equation}
0 = \left[ -\delta(v_0)- \frac{\hbar^2}{2m^\star} \nabla^2 + i\frac{\hbar}{2}\boldsymbol{\nabla}\cdot\boldsymbol{v}_0 + i\hbar\boldsymbol{v}_0\cdot\boldsymbol{\nabla} +\hbar gn_0 + \hbar g_\mathrm{res} n_\mathrm{res} - i\frac{\hbar \gamma}{2} \right] \sqrt{n_0} + \hbar|F_\tp|,    
\end{equation}
for the stationary part of the field, with $\gamma = \gamma_{\mathrm{LP}} + \gamma_\mathrm{in}$ and where the effective detuning is defined as $\delta(v_0) = \omega_\tp - \omega_0 - \frac{m^\star \boldsymbol{v}_0^2}{2\hbar}$

To study the excitations, we consider small perturbations around the stationary state (we neglect the perturbations of the reservoir density field~\cite{amelio_galilean_2020})
\begin{equation}
\psi(\boldsymbol{r},t) = e^{i(\phi_0(\boldsymbol{r}) - \omega_\tp t)} \left( \sqrt{n_0} + \delta\psi\, e^{-\gamma t/2} \right) = \psi_0 + \psi_1,
\end{equation}
where $\psi_1 = e^{i(\phi_0(\boldsymbol{r}) - \omega_\tp t)} e^{-\gamma t/2} \delta\psi$. Inserting this into Eq.~\eqref{eq:dGPEannex} and linearizing to first order in $\psi_1$ yields the Bogoliubov-de Gennes equations.

The interaction term becomes
$$
\hbar g|\psi|^2 \psi \approx \hbar g \left( |\psi_0|^2 \psi_0 + 2|\psi_0|^2 \psi_1 + \psi_0^2 \psi_1^* \right),
$$
and the kinetic energy term
$$
-\frac{\hbar^2}{2m^\star} \nabla^2 \psi 
\approx 
-\frac{\hbar^2}{2m^\star}
\nabla^2 \psi_0 
-\frac{\hbar^2}{2m^\star} \left( \boldsymbol{\nabla}^2 + 2i\, \boldsymbol{\nabla} \phi_0(\boldsymbol{r}) \cdot \boldsymbol{\nabla} + i\, \boldsymbol{\nabla}\cdot \boldsymbol{\nabla} \phi_0(\boldsymbol{r}) \right)\psi_1.
$$

Using the fact that $\psi_0$ satisfies \eqref{eq:dGPEannex}, $\boldsymbol{v}_0 = \hbar\boldsymbol{\nabla}\phi_0/m^{\star}$ and writing $\psi_1 = \delta\psi\, e^{i(\phi_0(\boldsymbol{r}) - \omega_\tp t)}$, we obtain:

\begin{align}
i\hbar \partial_t \delta\psi &= \Big[ -\hbar\delta(v_0) - \frac{\hbar^2}{2m^\star} \nabla^2 - i\hbar \boldsymbol{v}_0 \cdot \nabla - i\frac{\hbar}{2} \nabla \cdot \boldsymbol{v}_0 + 2\hbar g n_0 + \hbar g_\mathrm{res} n_\mathrm{res}\Big] \delta\psi + \hbar g n_0 \delta\psi^*, \\
i\hbar \partial_t \delta\psi^* &= -\Big[ -\hbar\delta(v_0) - \frac{\hbar^2}{2m^\star} \nabla^2 + i\hbar \boldsymbol{v}_0 \cdot \nabla + i\frac{\hbar}{2} \nabla \cdot \boldsymbol{v}_0 + 2\hbar g n_0 + \hbar g_\mathrm{res} n_\mathrm{res} \Big] \delta\psi^* - \hbar g n_0 \delta\psi.
\end{align} These equations can be recast in matrix form
$$
i\hbar \partial_t 
\begin{pmatrix}
\delta\psi \\
\delta\psi^*
\end{pmatrix}
=
\begin{bmatrix}
-\hbar\delta(v_0) + 2\hbar g n_0 + \hbar g_{\mathrm{res}} n_{\mathrm{res}} + D & \hbar g n_0 \\
-\hbar g n_0 & -(-\hbar\delta(v_0) + 2\hbar g n_0 + \hbar g_{\mathrm{res}} n_{\mathrm{res}} + D^\star)
\end{bmatrix}
\begin{pmatrix}
\delta\psi \\
\delta\psi^*
\end{pmatrix},
$$
where the differential operator is given by
$$
D = -\frac{\hbar^2}{2m^\star} \boldsymbol{\nabla}^2 - i\hbar \boldsymbol{v}_0 \cdot \boldsymbol{\nabla} - i\frac{\hbar}{2} \boldsymbol{\nabla} \cdot \boldsymbol{v}_0.
$$
The first is the kinetic energy in the "fluid frame", the second is the Doppler term due to the fluid motion.
The third term is purely imaginary and its sign is governed by the velocity divergence.
The presence of this term is unique to the driven dissipative quantum fluids. Indeed, in the conservative case, this term is absorbed in the chemical potential.  

\subsection{Vortex case}
In the vortex configuration, the stationary solution takes the form $\psi_C(\boldsymbol{r}, t) = f(r) e^{i(C\theta + S(r) - \omega_\tp t)}$.
Since the system remains invariant under time translation and azimuthal rotation, the eigenmodes are plane-waves of the azimuthal angle
$$
\delta\psi(r, \theta, t) = u(r) e^{i(m\theta - \omega t)} + v^*(r) e^{-i(m\theta - \omega t)}.
$$
Introducing the radial Bogoliubov spinor $\ket{\delta\psi} = (u(r), v(r))^T$, the Bogoliubov dynamics reduces to the eigenvalue problem
\begin{equation}
(\hbar\omega - \mathcal{L}_{C, m}) \ket{\delta\psi} = 0,
\label{bdg_vortex}    
\end{equation}
with 
\begin{equation}
\mathcal{L}_{C, m} =
\begin{bmatrix} 
-\hbar\delta(v_0) + 2\hbar g n_0 + \hbar g_{\mathrm{res}} n_{\mathrm{res}} + D_+ & \hbar g n_0 \\
-\hbar g n_0 & -(-\hbar\delta(v_0) + 2\hbar g n_0 + \hbar g_{\mathrm{res}} n_{\mathrm{res}} + D_{-}^*)    
\end{bmatrix},
\end{equation}

where the differential operator acting on the radial coordinate writes
$$
D_\pm = -\frac{\hbar^2}{2m^\star}\left(\partial_r^2+\frac{1}{r}\partial_r\right) + \frac{\hbar^2 m^2}{2m^\star r^2} +\left(\pm \frac{\hbar^2 m C}{m^\star r^2} - \ci\hbar v_r \partial_r \right) - \frac{i\hbar}{2r}\partial_r (rv_r).
$$

\subsection{WKB approach}
As the equation Eq.~\eqref{bdg_vortex} is not analytically solvable, numerical resolution or using approximations are the only way to go.
The standard analytical procedure for studying the dynamics is the approximation Wentzel-Kramers-Brillouin (WKB)\cite{berry_semiclassical_1972} approximation.
This consists in writing the Bogoliubov excitations as a field with a fast-varying phase and a slowly varying envelope.
The Bogoliubov eigenmodes can thus be locally approximated as radial plane waves $\delta\psi_{\mathrm{WKB}}(r, \theta, t) = A(r)\ee^{i\int p(r)r}e^{\ci\left(m\theta-\omega t\right)}+B(r)\ee^{-i\int p(r)r}\ee^{-\ci\left(m\theta-\omega t\right)}$ with a local wave vector $(p(r), m/r)$~\cite{patrick_quantum_2022,delhom_entanglement_2024} with the following local relation dispersion obtained by replacing the mean field values in Eq.~\eqref{eq:lfdisp} by local ones
\begin{equation}
      \label{eq:wkbannex}
      \omega_\mathrm{B}^\pm\left(p, \frac{m}{r}\right)=\frac{\hbar}{m^{\star}}\frac{mC}{r^2} +v_rp \pm \sqrt{\left(\frac{\hbar}{2m^\star}\left(\left(\frac{m}{r}\right)^2+p^2\right) -\delta(v) + 2gn + g_\mathrm{res}n_\mathrm{res} \right)^2 - (gn)^2}
\end{equation}.

\section{Speed of sound and ergosurface definition}\label{app:soundspeed}

Under resonant pumping, when the pump frequency and wavevector are chosen such that $\delta(v) = g_\mathrm{res}n_\mathrm{res} + gn$, the Bogoliubov excitation spectrum in the WKB approximation ~\eqref{eq:wkbannex} reduces to
\begin{equation}
\omega_\mathrm{B}^{\pm}(\boldsymbol{k}) = \boldsymbol{v}(r)\cdot\,\boldsymbol{k}(r) \pm \sqrt{ \frac{\hbar k(r)^2}{2m^{\star}} \left( \frac{\hbar k(r)^2}{2m^{\star}} + 2g n(r) \right) },
\label{eq:TPbogo}
\end{equation}
which corresponds to the standard Bogoliubov dispersion of the GPE in the absence of pumping and losses.
At low wavevector, $\omega_B^\pm \underset{k\to 0}{\sim} \boldsymbol{v}\cdot\boldsymbol{k} \pm c_s |k|$, with $c_s = \sqrt{\hbar gn / m^\star}$ the speed of sound in the polariton fluid. This defines the hydrodynamic regime. In this limit, negative energy excitations ($\omega_\mathrm{B}^+ < 0$) arise when $|v(r)| > c_s$, which is the usual criterion for their onset. In rotating systems, the equation $|v(r)| = c_s$ defines the ergosurface $r_{\mathrm{erg}}$, separating an inner ergoregion ($|v(r)| > c_s$) containing negative energy modes from an outer region where all excitations have positive energy.

However, if $\delta(v) < g_\mathrm{res}n_\mathrm{res} + gn$, the linear dispersion of low wave vectors is lost, so the notion of a hydrodynamic speed of sound does not make sense anymore and the definition $|v(r)| = c_s$ for the ergosurface no longer holds. However, the notion of an ergoregion remains robust and can still be defined from condition $\omega_\mathrm{B}^+ < 0$.

In this case, the Bogoliubov dispersion \eqref{eq:wkbannex} can be rewritten as
\begin{align}
    \label{eq:lfbogom}
    \omega^\pm_\mathrm{B}(\boldsymbol{k}) &= \boldsymbol{v}(r)\cdot\boldsymbol{k}(r) \nonumber\\
    &\quad \pm \Biggl[ \left( \frac{\hbar k(r)^2}{2m^\star} \right)^2 
    + \frac{\hbar k(r)^2}{m^\star} \left( 2gn - \delta(v(r)) + g_\mathrm{res}n_\mathrm{res} \right) \nonumber \\
    &\quad\quad + \left( gn - \delta(v(r)) + g_\mathrm{res}n_\mathrm{res} \right) \left( 3gn - \delta(v(r)) + g_\mathrm{res}n_\mathrm{res} \right) \Biggr]^{1/2} - \ci\frac{\gamma}{2} \nonumber \\
    &= \boldsymbol{v}(r)\cdot\boldsymbol{k}(r) 
    \pm \sqrt{ \left( \frac{\hbar k(r)^2}{2m^\star} \right)^2 
    + c_\mathrm{B}^2 \left( k(r)^2 + \frac{m_\mathrm{det}}{\hbar^2} \right) },
\end{align}
where
\begin{equation}
\label{eq:csandmdet}
\begin{split}
    c_\mathrm{B} &= \sqrt{ \frac{ \hbar \left( 2gn - \delta(v) + g_\mathrm{res}n_\mathrm{res} \right) }{m^\star} }, \\
    m_\mathrm{det} &= m^\star \frac{ \sqrt{ \left( gn - \delta(v) + g_\mathrm{res}n_\mathrm{res} \right) \left( 3gn - \delta(v) + g_\mathrm{res}n_\mathrm{res} \right) } }{ 2gn - \delta(v) + g_\mathrm{res}n_\mathrm{res} }.
\end{split}
\end{equation}
The dispersion relation becomes analogous to that of a relativistic scalar field with mass $m_\mathrm{det}$ and light cone velocity $c_\mathrm{B}$, which sets an upper bound on the group velocities of the excitations~\cite{falque_polariton_2024}.

Although an analytical criterion for the location of the ergoregion becomes more complex in this regime, one can resort to numerical analysis. It can be observed that condition $\omega^+_\mathrm{B}(k) < 0$ becomes stricter than the linear case condition ($|v(r)| > c_s$), and the velocity on the ergosurface always satisfies $v(r_{\mathrm{erg}}) > \sqrt{\hbar gn/m^\star}$.

\section{Measurement of the interaction term}\label{SM_den}

We know that the photonic field resulting from cavity losses is proportional to the polariton field, so the photonic intensity is proportional to the interaction term $gn$. Since the photonic field is spatially resolved on a CMOS camera, $gn(\boldsymbol{r})$ can be accessed by calibrating the signal read by the camera.
This section details the procedure used to perform this calibration.

\begin{figure}[ht!]
    \centering
\includegraphics[width=1\textwidth]{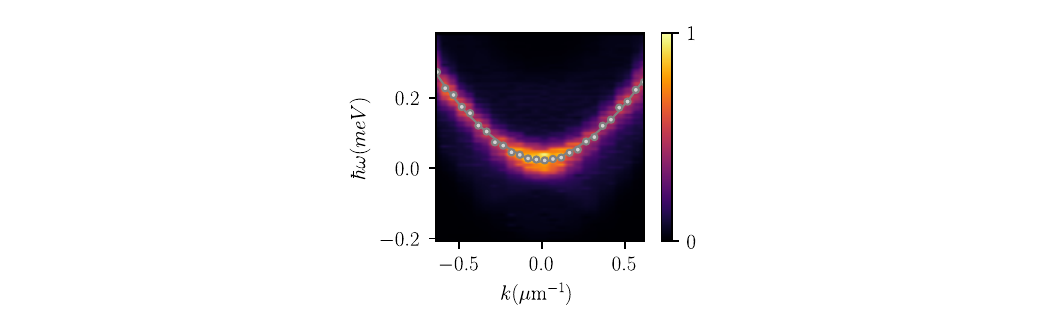}
    \caption{($\hbar\omega-k$) probe transmission intensity map on top of a polariton field excited by a gaussian beam at $\boldsymbol{k}_\tp=0$ . Each $k$ slice is fitted with a Lorentzian. The circled gray dots show the centers of these Lorentzian functions. The Lorentzian centers are then fitted by Eq.~\ref{eq:lfdisp} with $gn$ and $g_\mathrm{res}n_\mathrm{res}$ as free parameters, plotted in gray.
    }
    \label{fig:SM_reservoir}
\end{figure}

As mentioned previously, the dynamics of the polariton field coupled to a dark excitonic reservoir is governed by the following set of coupled equations~\eqref{eq:dGPEannex}, from which we see that $n_{\mathrm{res}}/n = \gamma_{\mathrm{in}}/\gamma_{\mathrm{res}}$.
By fitting a probe transmission dispersion (see Fig.~\ref{fig:SM_reservoir}) on top of a Gaussian pump at $\boldsymbol{k}_\tp=0$ using Eq.~\ref{eq:wkbannex}, it is possible to find the ratio $\beta = \frac{g_\mathrm{res}n_\mathrm{res}}{gn} = \frac{g_{\mathrm{res}}\gamma_{\mathrm{in}}}{g\gamma_r}$. We obtain $\beta = 0.49\pm0.03$.

\begin{figure}[ht!]
    \centering
\includegraphics{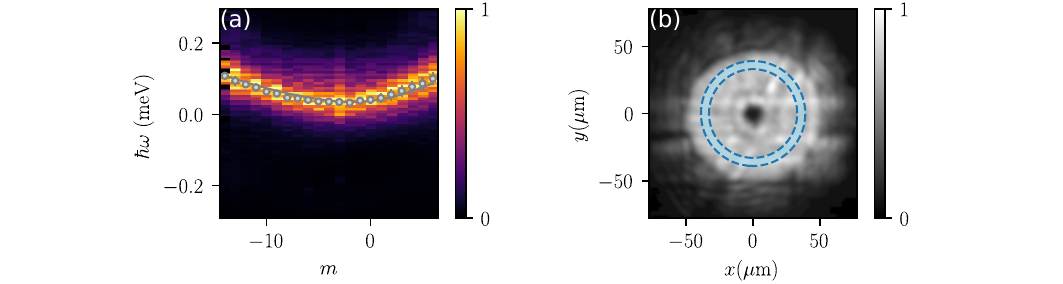}
    \caption{\textbf{Probe transmission over a vortex}.
    \textbf{a} $(\hbar\omega-m)$ probe transmission intensity map on top of a $C=4$ vortex. The transmission is filtered numerically for $r\in [\SI{33}{\micro\meter};\SI{39}{\micro\meter}]$. Each $k$ slice is fitted with a Lorentzian. The gray circled dots show the centers of these Lorentzian functions. The Lorentzian centers are then fitted by Eq.~\eqref{eq:lfdisp} with $gn$ as free parameter, plotted in gray.
    \textbf{(b)} Density map of the $C=4$ vortex field. The blue area is the filtered location to compute the transmission of \textbf{(a)}.}
    \label{fig:SM_gn}
\end{figure}

Once this value is known, the transmission of the probe on top of the vortex fluid is fitted by Eq.~\ref{eq:lfdisp} for $r = r_\mathrm{fit} = \SI{36}{\micro\meter}$ (far from regions where the field gradient is high, that is, the vortex core and outer part) fixing $g_\mathrm{res} n_\mathrm{res} = \beta\times gn$, see Fig.~\ref{fig:SM_gn}. This gives a $gn(r_{\mathrm{fit}})$ to calibrate the photonic signal recorded by the camera. As the photonic intensity is linear in loss intensity, we can use this calibration across the entire camera field to obtain the $gn$ map.

\section{OAI for Bogolioubov field measurements}

Even though the pump and the probe are crossed-polarized, due to their strong intensity imbalance, there are still residual light coming from the pump. OAI is a powerful technique that allows temporal separation between the pump and the probe. This makes our adaptation of coherent probe spectroscopy~\cite{claude_high-resolution_2022} particularly well suited to situations where the pump and probe modes do not separate spatially during propagation, as is the case here. It can also be used to resolve the phase of the Bogoliubov field, that was inaccessible with the previous implementation~\cite{claude_high-resolution_2022}.

Both the pump and probe lasers are sent into the cavity and imaged together with a reference probe pick-up before the SLM. This reference beam has a large waist, is flat in phase, and is set to a controlled in-plane wavevector $\boldsymbol{k}_\mathrm{ref}$ so that it can be approximated by a plane wave $Re^{\ci (\omega_\mathrm{pr}t-\boldsymbol{k}_{\mathrm{ref}}\boldsymbol{r})}$. 

The signal acquired by the camera is
\begin{align*}
\mathrm{CMOS}_{\mathrm{den}} &\propto \left\langle \left| f(\boldsymbol{r}) + u(\boldsymbol{r})\, \mathrm{e}^{-\ci\omega t} + v^*(\boldsymbol{r})\, \mathrm{e}^{i \omega t} + R(\boldsymbol{r})\, \mathrm{e}^{\ci\left(\boldsymbol{k}_{\mathrm{ref}} \cdot \boldsymbol{r} -\omega t\right)} \right|^2 \right\rangle_{t_{\mathrm{exp}}} \\
\mathrm{CMOS}_{\mathrm{den}} &\propto \left\langle |f|^2 + |u|^2 + |v|^2 + R^2 \right. \\
&\quad + 2R|u|\cos(\boldsymbol{k}_{\mathrm{ref}} \cdot \boldsymbol{r} - \arg(u)) \\
&\quad + 2R|v|\cos(\boldsymbol{k}_{\mathrm{ref}} \cdot \boldsymbol{r} - \arg(v) - 2\omega) \\
&\quad \left. + 2R|f|\cos(\boldsymbol{k}_{\mathrm{ref}} \cdot \boldsymbol{r} - \arg(f) - \omega) \right\rangle_{t_{\mathrm{exp}}},
\end{align*}
where $\omega = \omega_{\mathrm{pr}} - \omega_{\mathrm{p}}$.  
Because $\omega \gg 1 / t_{\mathrm{exp}}$, the last two terms vanish after time averaging, so
\begin{align*}
\mathrm{CMOS}_{\mathrm{den}} &\propto |f|^2 + |u|^2 + |v|^2 + R^2 + 2R|u|\cos(\boldsymbol{k}_{\mathrm{ref}} \cdot \boldsymbol{r} - \arg(u)) \\
&\propto B + R u\, \mathrm{e}^{-i \boldsymbol{k}_{\mathrm{ref}} \cdot \boldsymbol{r}} + \text{c.c.},
\end{align*}
where $B(\boldsymbol{r}) = |f(\boldsymbol{r})|^2 + |u(\boldsymbol{r})|^2 + |v(\boldsymbol{r})|^2 + R(\boldsymbol{r})^2$.

To recover the signal of interest we need to demodulate it, so we take the Fourier transform.
\begin{align*}
\mathrm{TF}(\mathrm{CMOS}_{\mathrm{den}}) &\propto \tilde{B}(\boldsymbol{k}) + \mathrm{TF}(Ru) * \delta(\boldsymbol{k} - \boldsymbol{k}_{\mathrm{ref}}) + \mathrm{TF}(Ru^{\star}) * \delta(\boldsymbol{k} + \boldsymbol{k}_{\mathrm{ref}}) \\
&\propto \tilde{B}(\boldsymbol{k}) + \mathrm{TF}(Ru)(\boldsymbol{k} - \boldsymbol{k}_{\mathrm{ref}}) + \mathrm{TF}(Ru^{\star})(\boldsymbol{k} + \boldsymbol{k}_{\mathrm{ref}}).
\end{align*}

\begin{figure}[!ht]
    \centering
\includegraphics[width=\textwidth]{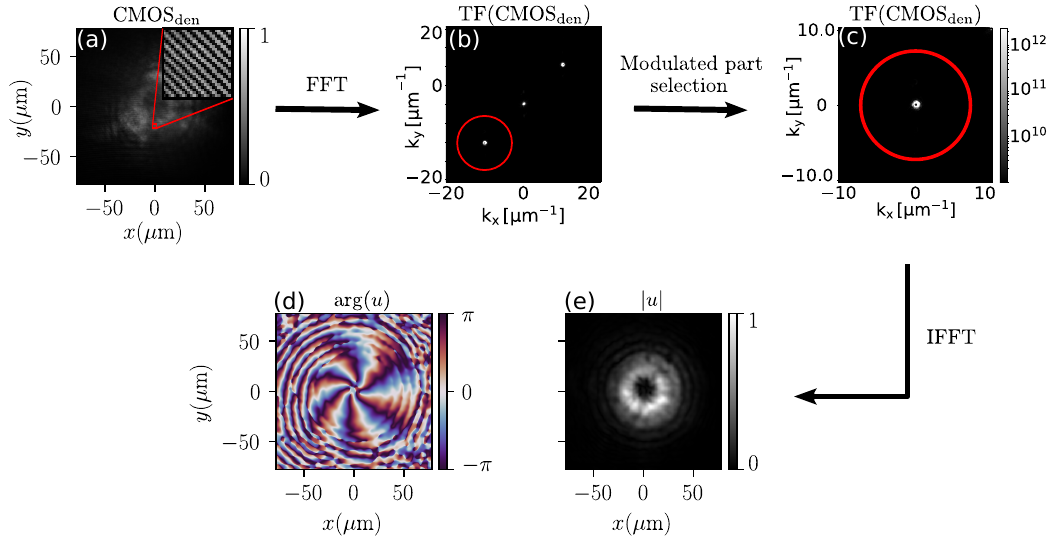}
    \caption{\textbf{OAI Fourier analysis of a probe mode}.
    \textbf{(a)} Interferogram ($\mathrm{CMOS}_{\mathrm{den}}$) acquired by the camera.
    \textbf{(b)} Fourier space intensity of the interferogram obtain by Fourier transform of \textbf{(a)}. The modulated part, where the signal of interest lies, is circled in red.
    \textbf{(c)} Region of interest of \textbf{(b)} containing the signal.
    \textbf{(d)~(e)} Phase and amplitude of the probe beam component $u$ extracted by Inverse Fourier transform after selection of the modulated part in Fourier space. 
    }
    \label{fig:SM_OAI}
\end{figure}

Now because $\boldsymbol{k}_{\mathrm{ref}}$ is much larger than the typical width of $\mathrm{TF}(Ru)$ (see Fig.~\ref{fig:SM_OAI}(b)), these three Fourier components can be separated.  
We then have access to $R(\boldsymbol{r})u(\boldsymbol{r})$ by the inverse Fourier transform and $u$ can be obtained by dividing by $R$, which can be obtained from an independent measurement.

By measuring $|f|^2$ independently, it is possible to get the intensity of the negative frequency component $|v|^2$. However, the phase reconstruction of this component is not possible with this technique.   
%\bibliography{biblio}
% \bibliography{biblio}